\documentclass[prb,twocolumn,showpacs,groupedaddress,amsmath,amssymb]{revtex4-1}

\usepackage{graphicx}
\usepackage{tabularx}
\usepackage{soul}

\begin{document}


\title{First-principles study of multi-control graphene doping using light-switching molecules}


\author {J.P. Trinastic}
\author {Hai-Ping Cheng}
 \affiliation {Department of Physics and Quantum Theory Project, University of Florida, Gainesville, Florida, 32611, USA}


\date{\today}


\begin{abstract}
 The high carrier mobility in graphene promises its utility in electronics applications.  Azobenzene is a widely studied organic molecule for switchable optoelectronic devices that can be synthesized with a wide variety of ligands and deposited on graphene.  Using first-principles calculations, we investigate graphene doping by physisorbed azobenzene molecules with various electron-donating and $-$accepting ligands.  We confirm previous experimental results that demonstrate greater p-doping of graphene for the $\textit{trans}$ compared to $\textit{cis}$ configuration when using a SO$_3$ electron-accepting ligand, however we find that NO$_2$ ligands maximize the p-doping difference between isomers.  We also examine how these doping effects change when the graphene monolayer is supported on a silica substrate.  We then extend these findings by examining the doping effects of an applied electrical bias and mechanical strain to the graphene, which lead to changes in doping for both the $\textit{trans}$ and $\textit{cis}$ isomers.  These results demonstrate a new type of multi-control device combining light, electric field, and strain to change carrier concentration in graphene.
\end{abstract}


\maketitle



\begin{center}\section{Introduction}\end{center}
\label{Introduction}

Graphene, the remarkable two-dimensional sheet of carbon atoms arranged in a honeycomb lattice, has demonstrated charge mobility greater than 10$^4$ cm$^2$V$^{-1}$s$^{-1}$ and charge carriers with relativistic speeds of 10$^6$ ms$^{-1}$.~\cite{novoselov2005two,du2008approaching}.  These fundamental properties of graphene make it a promising material for the electronics applications, however, for these applications to be realized, new methods to control carrier concentration must be determined.  Both experimental and first-principles studies have shown that carrier concentration in graphene can be modulated by voltage gating~\cite{pisana2007breakdown,Novoselov22102004,novoselov2005two,PhysRevLett.98.166802}, substitutional doping~\cite{wei_ndoped_graphene,B_N_graph_doping,chem_graph_overview}, or charge transfer doping from adsorbed atoms,  molecules, and clusters~\cite{chem_graph_overview,i2_br2_graph_doping,PhysRevB.82.075422,PhysRevB.82.245423,k_rb_gr_doping,aromatic_gr_doping,mol_gr_doping,PhysRevB.85.165444}.  

Substitutional doping, usually achieved by replacing carbon with boron or nitrogen atoms in the honeycomb lattice~\cite{wei_ndoped_graphene,B_N_graph_doping}, can have detrimental effects on graphene's intrinsically high mobility due to disruption of the $sp^2$ hybridization.  In contrast, molecular doping leaves the graphene lattice relatively intact, preserving graphene's intrinsic properties while also promoting charge transfer to tune the carrier concentration.  This is especially true for physisorbed molecules on graphene, in which van der Waals interactions dominate the molecular binding and the Dirac point and band structure close to the Fermi level are preserved.      In this case, if charge transfers from the adsorbed molecule to the graphene, the Dirac point will shift to an energy below the Fermi level of pristine graphene, known as n-doping.  This will lead to electrons being the majority charge carrier in graphene. Conversely, charge transferred from graphene to the molecule will shift the Dirac point above the Fermi level of pristine graphene, known as p-doping.  In this case, holes will be the majority charge carrier.  Often, physisorbed molecules are anchored to the graphene surface using additional molecular linkers, such as pyrene, to increase binding energies~\cite{kim2011light}.

In addition to preserving graphene's intrinsic electronic properties, physisorbed organic molecules offer unique methods of doping that could lead to novel optoelectronic devices.  One such molecule is azobenzene (AB), which consists of two benzene rings connected by a dinitrogen linker.  In its ground state, known as the $\textit{trans}$ isomer, the two benzene rings are parallel to one another.  Upon UV illumination, the AB molecule isomerizes to the $\textit{cis}$ isomer either through inversion or rotation around the N=N bond such that the two benzene rings are out of plane with each other.  The $\textit{cis}$ isomer can isomerize back to $\textit{trans}$ either through thermal relaxation or exposure to visible light.

This fascinating property of AB has opened up many potential uses for photoswitching devices.  First-principles and experimental studies have demonstrated AB-based conductance switches~\cite{wang2012electronic,PhysRevLett.92.158301,PhysRevB.73.125445,del2007tuning} and mechanisms for optical storage ~\cite{hagen2001photoaddressable,kolpak2011azobenzene}.  Relevant to graphene, AB molecules functionalized with the SO$_3$Na electron-accepting ligand have been directly adsorbed on graphene and have shown reversible, photoswitchable p-doping upon UV illumination~\cite{Peimyoo2012_thickness,peimyoo2012}, indicating the potential for a graphene-based optoelectronic device.  According to these experimental results, the $\textit{trans}$ isomer binds to graphene such that the benzene rings are parallel to the graphene surface and demonstrates an induced hole concentration in graphene.  Upon UV illumination, the benzene ring functionalized with SO$_3$Na lifts off the graphene surface, increasing the distance between the SO$_3$Na ligand and the graphene surface and thus reducing charge transfer.  To ensure that the benzene ring with SO$_3$Na lifts off the surface in the experiment, the other benzene ring is functionalized with 2CH$_3$ to increase its binding to graphene.  Similar photoswitching results have also been found linking AB to a pyrene molecule adsorbed on graphene~\cite{kim2011light} as well as AB linked to graphene oxide hybrids~\cite{ab_graphene_oxide,zhang2010investigation}.

Despite these initially promising results, little is known both about the details of the chemistry at the AB-graphene interface or about the effects that different ligands functionalized to AB will have on doping levels.  In addition, no studies have investigated further methods of tuning the graphene doping in the presence of the AB molecule, such as voltage gating or mechanical strain to lead to multi-control graphene doping.  In this paper, we present first-principles data investigating graphene doping by AB using a range of possible electron-accepting and -donating ligands as well as investigate the effect of an applied electric potential and mechanical strain on doping.  We then compare doping of a graphene monolayer to graphene on an amorphous SiO$_2$ substrate to better model experimental conditions~\cite{peimyoo2012} and investigate substrate effects.

The paper is organized as follows.  Section II outlines the computational methods used throughout the paper.  Section III.A reports electronic strucure and charge transfer results due to AB molecular absorption on graphene.  Section III.B examines the effects of different ligands linked to AB on graphene doping.  Section III.C discusses the effect of an applied electric field or mechanical strain to the AB-graphene system to further modulate doping levels.  We then compare AB doping of a graphene monolayer to graphene on an amorphous SiO$_2$ substrate with and without dangling surface bonds in Section III.D.  Conclusions are discussed in Section IV.\newline

\begin{center}\textbf{II. METHODS}\end{center}
\label{Methods}

A 7x7 rhombus supercell is used that consists of a 98-atom graphene monolayer and one of the adsorbed molecules, each of which ranges between 20-35 atoms. Although a long list of possible ligands exist, here we choose a collection of the strongest electron-donating (NH$_2$, OH) and -accepting (SO$_3$, SO$_3$Na, NO$_2$, CF$_3$) ligands as the best potential candidates for doping.  We label AB molecules based on the ligand attached to each of the two benzene rings, e.g., SO$_3$-AB-NO$_2$ corresponds to an AB molecule with a SO$_3$ ligand linked to one benzene ring and a NO$_2$ ligand linked to the other ring.  We also include the weakly-donating ligand, CH$_3$, to compare with previous experimental work that deposited an SO$_3$Na-AB-2CH$_3$ molecule on graphene~\cite{peimyoo2012}.  A 15 $\textup{\AA}$ vacuum extends along the z-axis to prevent interactions between periodic images.  We complete all first-principles calculations of binding energies, charge densities, and density of states (DOS) using the Vienna $\textit{ab initio}$ simulation package (VASP)~\cite{blochl1994projector,kresse1999ultrasoft}, which uses a plane-wave basis set to solve the Kohn-Sham equations.  A 6x6x1 Monkhorst-Pack~\cite{monkhorst1976special} $\textit{k}$-point mesh is used for the Brillouin Zone (BZ) integration in all cases, with a 600 eV energy cutoff for the wavefunctions.  For DOS calculations, the $\textit{k}$-point mesh is increased to 24x24x1 and a 0.025 eV smearing is implemented.

All self-consistent supercell calculations are performed using the Perdew-Burke-Ernzerhof generalized gradient approximation (PBE-GGA) of density functional theory (DFT).  In order to account for the van der Waals (vdW) interactions present betwen graphene and adsorbates, we use the opt86-vdW functional implemented in VASP that includes nonlocal correlation responsible for the dispersion interactions ~\cite{PhysRevB.83.195131}.  All structures are relaxed until the force on each atom is less than 0.01 eV/$\textup{\AA}$.

To provide a holistic picture of charge transfer at the graphene-AB interface, we quantify charge transfer in two ways.  First, we compute the total charge on each atom using Bader charge analysis~\cite{tang2009grid} and calculate the difference in total charge on the graphene monolayer alone compared to graphene with each adsorbed molecule.  Second, we calculate the shift in energy of the Dirac point of graphene with each adsorbed molecule relative to its position for pristine graphene.  This provides a measure of n-doping (Dirac point shifted below) or p-doping (Dirac pointed shifted above) that can be related to graphene carrier concentration to compare with experiment.  Previous experimental results have demonstrated that, for small doping levels, the Dirac point energy shift $E_{DP}$ (measured relative to the Fermi level) and carrier concentration $n$ can be related by $E_{DP} = \hbar|v_F|\sqrt{\pi n}$, where $v_F$ is the Fermi velocity~\cite{novoselov2005two}.  Therefore, for each AB derivative investigated, we report binding energy (BE), charge transfer (CT), Dirac point shift (DPS), and carrier concentration (CC) calculated from the DPS using the previous equation.  Across all molecules tested, we find that a magnitude of charge transfer less than $|$0.05$|$e never results in a shift in the Dirac point within the precision of our calculations.  Therefore, we will only consider ligands that result in significant charge transfer greater than $|$0.05$|$e as viable for significant graphene doping.

We also investigate the effect of electric bias and mechanical strain on AB-induced graphene doping (Section III.C).  To examine bias, we apply a homogeneous electric field along the z-axis (perpendicular to the graphene monolayer).  A linear electrostatic potential correction is added to account for errors due to the periodic boundary conditions in the calculation~\cite{neugebauer1992adsorbate}.  We vary the magnitude of the electric field from 0.0 to $\pm$0.5 V/$\textup{\AA}$.  To examine how mechanical strain impacts doping, we increase the unit cell size from 1 to 5 percent along either the armchair or zigzag edges of the graphene supercell.  We then re-relax the structure with the adsorbed molecule and calculate the CT and DPS.

\begin{figure*}
\includegraphics{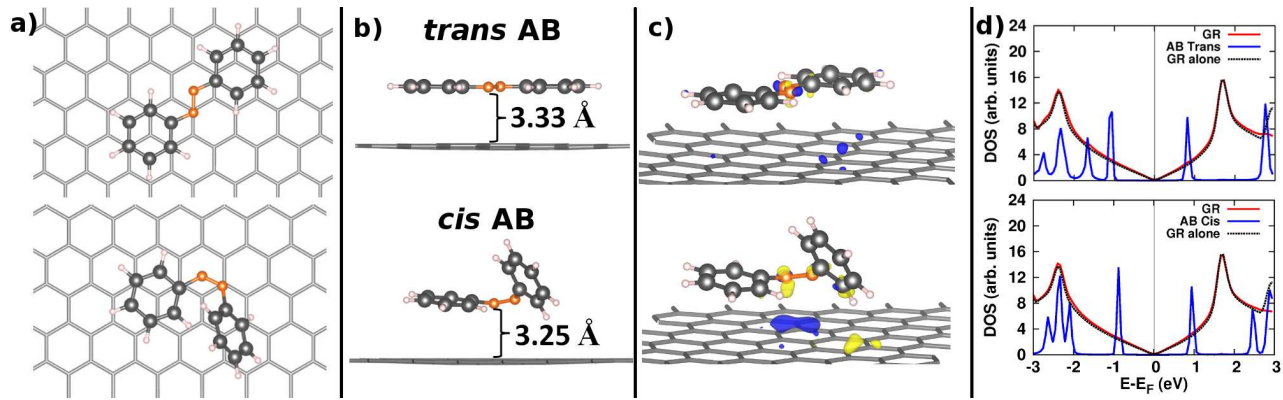}
\caption{\label{ab_image}(Color online) Geometric and electronic structure of azobenzene (AB) adsorbed on the graphene surface, including a) top view of optimized structure, b) side view of optimized structure, c) charge difference isosurfaces, where dark blue isosurfaces represent regions of hole accumulation and light yellow isosurfaces represent regions of electron accumulation, and d) partial density of states (PDOS).  The vertical black line in the PDOS plots represents the Fermi energy.  Upper panels refer to the $\textit{trans}$ isomer and bottom panels refer to the $\textit{cis}$ isomer.  The gray mesh corresponds to the graphene honeycomb lattice.  The large grey spheres are C atoms in the benzene ring, blue spheres are O atoms, orange spheres are N atoms, and the small white spheres are H atoms.}
\end{figure*}

After studying AB molecular doping of a graphene monolayer, we will investigate the effect of an amorphous SiO$_2$ substrate on doping levels (Section III.D).  For these calculations, we have used a 6x6 graphene supercell (72 C atoms).  To create the amorphous SiO$_2$ substrate of this size, we have used the LAMMPS molecular dynamics software~\cite{Plimpton19951} and the BKS SiO$_2$ potential~\cite{PhysRevLett.64.1955} to anneal a crystalline alpha-SiO$_2$ supercell containing 324 atoms that matches the experimental density (2.2 g/cm$^3$) with the same two-dimensional lattice constants as the graphene supercell.  The structure has been heated to 5000K, equilibrated, and then cooled to 300K over 120 ps.   After relaxing the structure at room temperature, we have cut the sample along an xy plane such that the sample is approximately 20 $\textup{\AA}$ along the z-axis, containing 240 atoms.  We have then run a final simluation at 100K to heal the cut surface and reduce the number of dangling bonds as much as possible.  We have then saturated any remaining dangling bonds by manually adding H atoms to unsaturated O atoms and OH molecules to unsaturated Si atoms.  After relaxing the final SiO$_2$ structure, the AB molecule and graphene monolayer have then been placed above the SiO$_2$ substrate and the structure has been relaxed using DFT.  Charge transfer to the molecule and the associated Dirac point shift have been calculated.  To simluate the presence of dangling bonds that lead to p-doping of graphene seen in experiment, we have compared these results to CT and DPS after removing one H atom at the SiO$_2$ surface.  We have created two different amorphous SiO$_2$ structures using the above procedure and averaged the CT and DPS results from each.

\begin{center}\textbf{III. RESULTS AND DISCUSSION}\end{center}
\label{results}

\begin{center}\textbf{A. Azobenzene Absorption on Graphene}\end{center}

Example optimized structures of the AB molecule physisorbed to the graphene surface are shown in Figure \ref{ab_image}(a)-(b).  Ligands are linked to AB by replacing the H atom farthest from the NN linker on either benzene ring with an electron-donating or -accepting ligand.  To isolate the effects of doping from the ligands alone, we first calculate the CT and DOS for the AB molecule without ligands adsorbed to the graphene surface.  After testing multiple adsorption sites including top (all benzene C atoms above graphene C atoms), between (half of benzene C atoms above C-C bonds in graphene), and hollow (half of benzene C atoms above the center of graphene honeycombs), we find that the stable $\textit{trans}$ configuration includes one benzene ring in the hollow position and the other slightly shifted from the hollow site, due to the difference in bond length between the AB NN linker (orange atoms in Figure \ref{ab_image}(a), upper panel) and the C-C distance (gray honeycomb lattice in Figure \ref{ab_image}(a)) of the underlying graphene substrate (1.28 vs 1.42 $\textup{\AA}$).  In the $\textit{cis}$ case, the benzene ring closest to the graphene surface is in the hollow position but tilted 11$^{\circ}$ from being parallel to the graphene substrate.  The second benzene ring is lifted from the surface and 58$^{\circ}$ tilted from parallel.

Respective binding energies are -1.26 eV for the $\textit{trans}$ and -0.91 eV for the $\textit{cis}$ isomer.  This corresponds to -0.053 eV/atom and -0.038 eV/atom, agreeing well with previous theoretical (-0.041 eV/atom) and experimental (-0.042 eV/atom) results for benzene adsorption on graphene or graphite.~\cite{PhysRevB.69.155406,PhysRevLett.96.146107}  The average AB-graphene distance, as measured by the perpendicular distance between one N linker atom in AB and the graphene surface, is 3.33 $\textup{\AA}$ for the $\textit{trans}$ and 3.25 $\textup{\AA}$ for the $\textit{cis}$ isomer (Figure \ref{ab_image}(b)).  These distances are expected since the dominant interactions between graphene and the benzene rings are similar to the $\pi$-$\pi$ stacking in graphite~\cite{aromatic_gr_doping}, and they do not significantly change when different ligands are attached to the AB molecule.  

\begin{table*}
\caption{\label{table_ligand}Ligand type, binding energy (BE), charge transfer (CT), Dirac point shift (DPS), and carrier concentration (CC) for azobenzene derivatives physisorbed to graphene that induce doping.  All values of CT indicate p-doping of graphene.  Values of DPS indicate the energy shift of the Dirac point above its position for pristine graphene.  Carrier concentration (CC) is calculated from the DPS (see equation in text).  For the $cis$ isomer, the molecule closest to the graphene is listed next to the isomer label.}
\begin{tabular}{l*{15}{c*2}r}
ligand type  & isomer & BE (eV) & CT (e) & DPS (eV) & CC ($\times$10$^{13}$) (cm$^{-2}$) & CC Ratio\\
\hline
no ligand & $trans$ & -1.26 & $<$$|$0.05$|$ & 0.00 & 0.00 \\
& $cis$ & -0.91 & $<$$|$0.05$|$ & 0.00 & 0.00 & 0.00 \\
SO$_3$-AB & $trans$  & -2.32 & 0.71 & 0.41 & 1.12 \\
 & $cis$-SO$_3$ & -1.88 & 0.69 & 0.41 & 1.12 & 1.00 \\
 & $cis$-none & -1.44 & 0.54 & 0.37 & 0.91 & 1.23 \\
SO$_3$-AB-2CH$_3$ & $trans$  & -2.37 & 0.60 & 0.40 & 1.07 \\
 & $cis$-SO$_3$ & -1.68 & 0.59 & 0.40  & 1.07 & 1.00 \\
 & $cis$-2CH$_3$ & -1.74 & 0.47 & 0.35  & 0.82 & 1.31\\
SO$_3$Na-AB-2CH$_3$ & $trans$  & -2.06 & $<$$|$0.05$|$ & 0.00 & 0.00 \\
 & $cis$-SO$_3$Na & -1.36 & $<$$|$0.05$|$ & 0.00  & 0.00 & 0.00 \\
 & $cis$-2CH$_3$ & -1.53 & $<$$|$0.05$|$ & 0.00 & 0.00 & 0.00 \\
SO$_3$-AB-SO$_3$ & $trans$ & -3.17 & 1.20 & 0.80 & 4.26 \\
& $cis$ & -2.45 & 0.99 & 0.63 & 2.64 & 1.61 \\
NO$_2$-AB-NO$_2$ & $trans$ & -1.70 & 0.13 & 0.25 & 0.42 \\
& $cis$ & -1.13 & 0.08 & 0.18 & 0.22 & 1.92 \\
NO$_2$-AB-CF$_3$ & $trans$  & -1.62 & 0.06 & 0.18 & 0.22  \\
 & $cis$-NO$_2$ & -1.12 & 0.06 & 0.17 & 0.19 & 1.12 \\
 & $cis$-CF$_3$ & -1.09 & 0.05 & 0.10 & 0.07 & 3.14 \\
SO$_3$-AB-CF$_3$ & $trans$  & -2.59 & 0.75 & 0.42 & 1.17  \\
 & $cis$-SO$_3$ & -2.08 & 0.75 & 0.42  & 1.17 & 1.00 \\
 & $cis$-CF$_3$ & -1.75 & 0.60 & 0.39 & 1.01 & 1.16 \\
SO$_3$-AB-NO$_2$ & $trans$  & -2.80 & 0.78 & 0.44 & 1.29 \\
 & $cis$-SO$_3$ & -2.15 & 0.78 & 0.43  & 1.23 & 1.05 \\
 & $cis$-NO$_2$ & -1.91& 0.69 & 0.41 & 1.12 & 1.15 \\
\hline
\end{tabular}
\end{table*}

For the isolated AB molecule without graphene, our calculations give dihedral CNNC and CCNN angles of 180.0$^{\circ}$ and 0.2$^{\circ}$ for the $\textit{trans}$ isomer and 11.1$^{\circ}$ and 52.5$^{\circ}$ for the $\textit{cis}$ isomer.  In addition, we find the NN and CN bond lengths to be 1.27 $\textup{\AA}$ and 1.42 $\textup{\AA}$ for $\textit{trans}$ and 1.25 $\textup{\AA}$ and 1.43 $\textup{\AA}$ for $\textit{cis}$. All these values agree well with previous DFT~\cite{tiago_ab_angle_bonds} and experimental~\cite{bouwstra1983structural,traetteberg1977gas} calculations and indicate that our functional provides accurate molecular geometries.  After binding to graphene, the $\textit{trans}$ dihedral CNNC and CCNN angles change only slightly to 179.3$^{\circ}$ and 1.4$^{\circ}$, respectively.  The $\textit{cis}$ dihedral angles change to 7.46$^{\circ}$ and 52.5$^{\circ}$, indicating that binding marginally rotates the upper benzene ring closer to graphene.  Nevertheless, the general geometry from the gas phase remains relatively intact, as expected for physisorption.

As shown in Figure \ref{ab_image}(c)-(d), neither the $\textit{trans}$ nor $\textit{cis}$ isomer significantly dope graphene as indicated by the identical graphene DOS with and without AB (Figure \ref{ab_image}(d), red and black lines, respectively).  Bader charge analysis indicates negligible charge transfer below the threshold of 0.05$|$e$|$ (Table I), matching the lack of a shift in the Dirac point.  The charge difference isosurfaces in Figure \ref{ab_image}(c) show minimal charge transfer for $\textit{trans}$; however, the $\textit{cis}$ isomer creates hole accumulation in graphene beneath the NN double bond and  electron accumulation below the lifted benzene ring.  These two localized pools roughly cancel to result in negligible net charge transfer below our threshold.

\begin{center}\textbf{B. Azobenzene Ligands and Graphene Doping}\end{center}

We next investigate the effects of ligands linked to the AB derivative on graphene doping.  For each ligand, we examine the binding energy (BE), charge transfer (CT), and Dirac point shift (DPS) due to the adsorbate.  As convention, all CT and DPS $>$ 0 represent charge transfer from graphene to the molecule and a shift of the Dirac point to higher energies (p-doping).  In the case of the $\textit{cis}$ isomer, if different ligands are attached to each benzene ring, two binding energies are calculated to account for the fact that either benzene ring can rotate during the isomerization from $trans$ to $cis$ and end up farther from the graphene surface.  This is especially important to consider when determinig optimal ligands to ensure that the ligand that contributes most to doping graphene lifts away from the surface upon photoisomerization to provide a doping difference between the $trans$ and $cis$ isomers.

We list in Table \ref{table_ligand} all ligand combinations that induce graphene doping as measured by both a nonzero DPS and CT $>$ $|$0.05$|$e.  Table 2 reports similar data for all ligands not inducing significant graphene doping.  Out of all possible ligands tested, SO$_3$-AB, SO$_3$-AB-2CH$_3$, SO$_3$-AB-SO$_3$, NO$_2$-AB-NO$_2$, NO$_2$-AB-CF$_3$, SO$_3$-AB-CF$_3$, and SO$_3$-AB-NO$_2$ provide significant charge transfer.  The common ligands across all these combinations are SO$_3$ and NO$_2$, indicating that open shell molecules are likely necessary to significantly dope graphene, as previous first-principles studies have also suggested~\cite{wehling2008molecular}.  Both of these ligands are strong electron acceptors, and therefore we only find evidence of p-doping across all ligands.  We also include results for SO$_3$Na-AB-2CH$_3$ because, in contrast to previous experiment~\cite{peimyoo2012}, we find that this ligand combination does not dope graphene, which we will discuss in more detail below.

Although all of the above ligand combinations demonstrate doping in both the $\textit{trans}$ and $\textit{cis}$ configurations, we must examine the binding energies to determine which will be effective for switching applications.  Binding energies are important because, in order to achieve a doping difference between the isomers, the ligand responsible for charge transfer must be lifted farther from the graphene surface in the $\textit{cis}$ isomer.  Therefore, that $\textit{cis}$ configuration should have a higher BE than the reverse configuration in which the electron-accepting ligand is closer to the graphene.  For example, examining the BE in Table I, we can see that for the SO$_3$-AB molecule, in which one benzene ring has a SO$_3$ ligand and the other has no ligand, the binding energy is 0.44 eV greater when the benzene ring with the SO$_3$ ligand is closer to the surface.  This means that, upon photoisomerization, the lone benzene ring will likely rotate away from the graphene surface.  But in this case, the CT and DPS are identical to the $\textit{trans}$ configuration, leading to an identical CC and a CC ratio of 1.00, indicating no switching behavior.  These results indicate that the ligand has a significant effect on binding to the graphene and must be chosen wisely to provie the correct switching behavior.

\begin{figure*}
\includegraphics{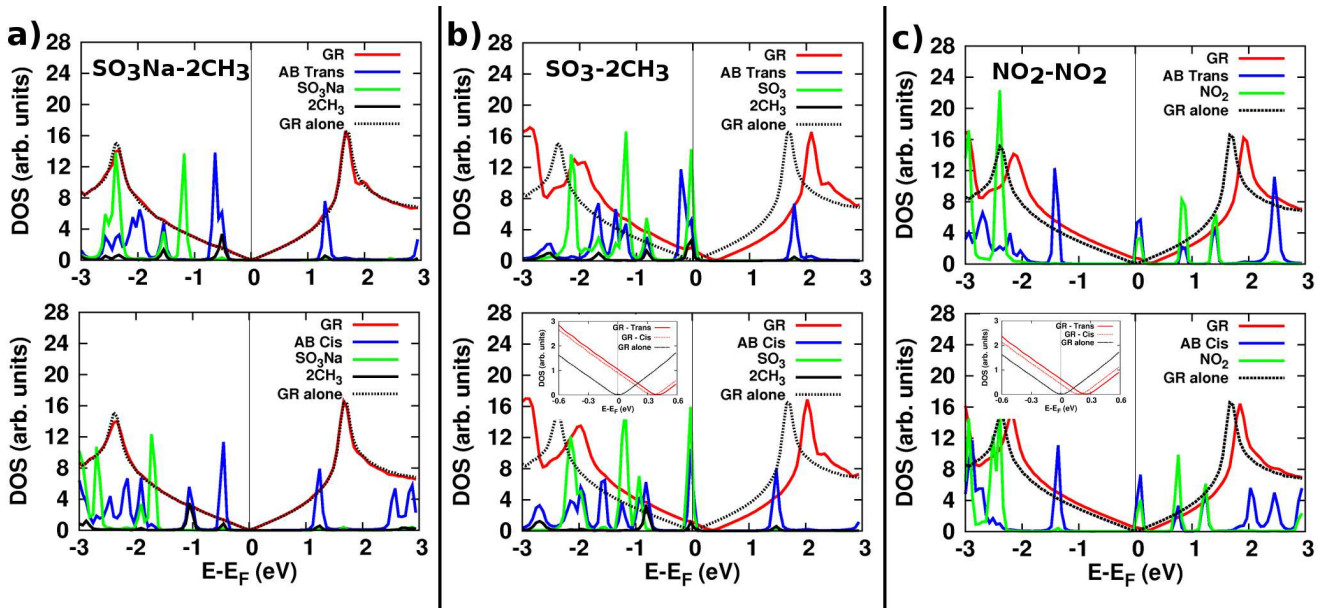}
\caption{\label{func_dos}(Color online) Projected density of states (PDOS) for a) SO$_3$Na-AB-2CH$_3$, b) SO$_3$-AB-2CH$_3$, and c) NO$_2$-AB-NO$_2$ adsorbed on graphene.  The vertical black line in each plot represents the Fermi energy.  Upper and lower panels are the PDOS for the $\textit{trans}$ and $\textit{cis}$ isomer, respectively, for each ligand type.  Plot insets in $\textit{cis}$ SO$_3$-AB-2CH$_3$ and NO$_2$-AB-NO$_2$ PDOS compare the graphene DOS between the $\textit{trans}$, $\textit{cis}$, or no isomer to emphasize the change in doping level across the three configurations.}
\end{figure*}

Enforcing this criterion, we narrow our choices for optimal ligands to SO$_3$-AB-2CH$_3$, SO$_3$-AB-SO$_3$, and NO$_2$-AB-NO$_2$ that demonstrate the largest CC ratios in which an electron-accepting ligand will lift away from graphene in the $cis$ configuration. Among the three remaining ligand combinations, NO$_2$-AB-NO$_2$ provides the greatest carrier concentration difference between the $\textit{trans}$ and $\textit{cis}$ isomers, with a CC ratio of 1.92, even though the CT and DPS are smaller compared to the SO$_3$ ligand.  In general, using the same electron-accepting ligand on both benzene rings should be ideal for switching applications such that either benzene ring can photoisomerize and lead to a change in graphene doping.  Hence, our calculations indicate that NO$_2$-AB-NO$_2$ is the optimal molecule to maximize doping differences between the $\textit{trans}$ and $\textit{cis}$ isomers.  These results will be valuable to guide future experimental study of AB switching behavior on graphene.

\begin{figure*}
\includegraphics{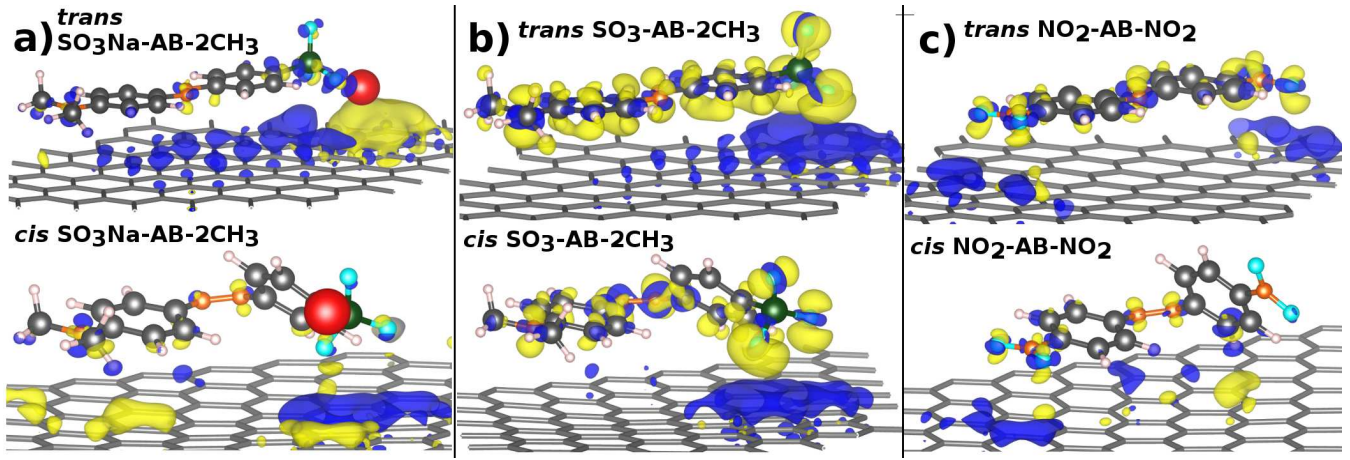}
\caption{\label{ch_density}(Color online) Charge difference isosurfaces for a) SO$_3$Na-AB-2CH$_3$, b) SO$_3$-AB-2CH$_3$, and c) NO$_2$-AB-NO$_2$.  Upper and lower panels are the $\textit{trans}$ and $\textit{cis}$ isomers, respectively, for each ligand type.  Dark blue isosurfaces represent regions of hole accumulation and light yellow isosurfaces represent regions of electron accumulation compared to the graphene or molecule alone. Medium grey spheres are C atoms, small white spheres are H atoms, medium orange are N atoms, medium light blue atoms are O atoms, large dark green atoms are S atoms, and the large red sphere is the Na atom.}
\end{figure*}

We next examine the projected density of states (PDOS) (Figure \ref{func_dos}) and charge difference isosurfaces (Figure \ref{ch_density}) of three ligand types representing various levels of graphene doping.  The upper and lower panels refer to the $\textit{trans}$ and $\textit{cis}$ cases for each ligand, respectively.  In Figure \ref{func_dos}(a), the PDOS for SO$_3$Na-AB-2CH$_3$ demonstrates a lack of graphene doping, as there is no shift in the Dirac point for graphene DOS with an adsorbed molecule (red line).  In addition, molecular AB and CH$_3$ states are 0.6 eV below the Fermi level and 1.3 eV above the Fermi level, indicating that electrons and holes in graphene could not thermally excite to either state.  Although this ligand combination has been reported in experiment to show differential doping between the $\textit{trans}$ and $\textit{cis}$ isomers~\cite{peimyoo2012}, we find that the graphene p-doping only occurs when we remove the Na and use a SO$_3$-AB-2CH$_3$ ligand combination, as shown in Figure \ref{func_dos}(b).  In this case, the $trans$ isomer leads to the Dirac point shifting 0.40 eV above the Fermi level, indicating p-doping, and the molecular orbitals corresponding to AB, the SO$_3$ ligand, and the 2CH$_3$ ligands are pinned just below the Fermi level.  This indicates that the transferred charge is delocalized across the SO$_3$ ligand and benzene rings, and, to a lesser extent, the CH$_3$ ligands.  For the $cis$ isomer, the Dirac point is 0.35 eV above the Fermi level, indicating decreased p-doping, and the same molecular states are pinned at the Fermi level.  The NO$_2$-AB-NO$_2$ molecule, shown in Figure \ref{func_dos}(c), also induces p-doping, as the $trans$ and $cis$ isomers shift the Dirac point 0.25 eV and 0.18 eV above the Fermi level, respectively.  In addition, molecular states corresponding to the AB molecule and NO$_2$ ligands are pinned just above the Fermi level (see Figure \ref{func_dos}(c)), similar to using the SO$_3$ ligand, indicating that the transferred charge delocalizes across the entire molecule.  We can compare this to 

To understand the difference between using the SO$_3$Na and SO$_3$ ligands, we examine the charge difference isosurfaces of the two cases in Figures \ref{ch_density}(a) and (b).  In the case of SO$_3$Na-AB-2CH$_3$, the Na atom bonds with one of the O atoms on the SO$_3$ ligand.  We test several configurations with the Na atoms bonded to O atoms either close to or far from the graphene surface.  We find the lowest energy configuration for the $trans$ isomer results in the Na atom bonded to the O atom closest to the graphene surface, with the Na atom 2.65 $\textup{\AA}$ from the graphene (Figure \ref{ch_density}(a), upper panel).  In this configuration, as shown in Figure  \ref{ch_density}(a), the Na atom creates a localized region of electron accumulation in the graphene lattice (yellow isosurface). This charge comes from the graphene region directly below the AB molecule (blue isosurface).  However, very little charge is transferred to the AB molecule, leading to no net graphene doping.  For the $cis$ isomer, Na lifts farther from the surface and therefore decreases the amount of localized charge density in the graphene beneath it.  A small region of decreased charge density beneath SO$_3$ appears, however this is balanced with the increased charge density from Na and the 2CH$_3$ ligands that results in no net charge transfer.

In contrast, as shown in Figure \ref{ch_density}(b), removing the Na atom leads to a significant region of hole accumulation in graphene directly beneath the SO$_3$ ligand using the $trans$ isomer, inducing a 0.40 eV DPS and 1.07$\times$10$^{13}$ cm$^{-2}$ carrier concentration.  This transferred charge is delocalized across the entire AB molecule and ligand.  The pattern is similar for the $cis$ case, however the amount the of transferred charge is reduced, leading to a 0.35 eV DPS, 0.82 $\times$10$^{13}$ cm$^{-2}$ carrier concentration, and a 1.31 CC ratio.  These results clearly demonstrate that the SO$_3$ ligand causes the charge transfer, and that the presence of the Na atom negates this effect due to charge localization in the graphene.  Our reported carrier concentration for the $trans$ SO$_3$-AB-2CH$_3$ isomer is of the same order of magnitude as that found in experiment (5$\times$10$^{13}$ cm$^{-2}$).~\cite{peimyoo2012}  A higher carrier concentration is expected in the experiment because the graphene was deposited on a silica (SiO$_2$) substrate that further p-doped the graphene.  More importantly, the CC ratio we calculate (1.31) agrees well with the experimental ratio (1.25), indicating our first-principles calculations without the Na atom identify the major physical mechanism through which graphene doping changes through photoisomerization.  Since our results using the SO$_3$ ligand without Na best match experimental results, we believe that the experimental methodology, in which AB molecules began in solution and were drop cast on the graphene substrate, likely led to Na dissociation from the molecule.  The previous experimental results did not report molecular concentrations and therefore we cannot confirm that our molecular concentration (2.96$\times$10$^{18}$ cm$^{-2}$) exactly matches experimental values.

Comparing the charge difference isosurfaces between the SO$_3$-AB-2CH$_3$ and NO$_2$-AB-NO$_2$ molecules can also explain why the latter provides the best CC ratio.  The SO$_3$ ligand in the SO$_3$-AB-2CH$_3$ molecule is the strongest electron acceptor in this case, and therefore significant charge is transferred in both the $trans$ (0.60$|$e$|$, Figure \ref{ch_density}(b), upper panel) and $cis$ (0.47$|$e$|$, Figure \ref{ch_density}(b), lower panel) cases.  On the other hand, the $trans$ NO$_2$-AB-NO$_2$ molecule only moderately p-dopes graphene in the regions directly beneath each NO$_2$ ligand (0.13$|$e$|$, Figure \ref{ch_density}(c), upper panel).  Thus, when the molecule isomerizes to $cis$, one of the NO$_2$ lifts far enough away from graphene that it no longer p-dopes the graphene at all (Figure \ref{ch_density}(c), lower panel).  Only the other NO$_2$ ligand p-dopes graphene and the raised benzene ring even slightly n-dopes graphene.  This leads to a significantly decreased net p-doping (0.08$|$e$|$), and a greater CC ratio using the NO$_2$ ligands (1.92) compared to the SO$_3$-AB-2CH$_3$ (1.31) or SO$_3$-AB-SO$_3$ (1.61) ligands.

\begin{center}\textbf{C. Effect of Electric Bias and Mechanical Strain on Doping}\end{center}

\begin{figure}
\includegraphics{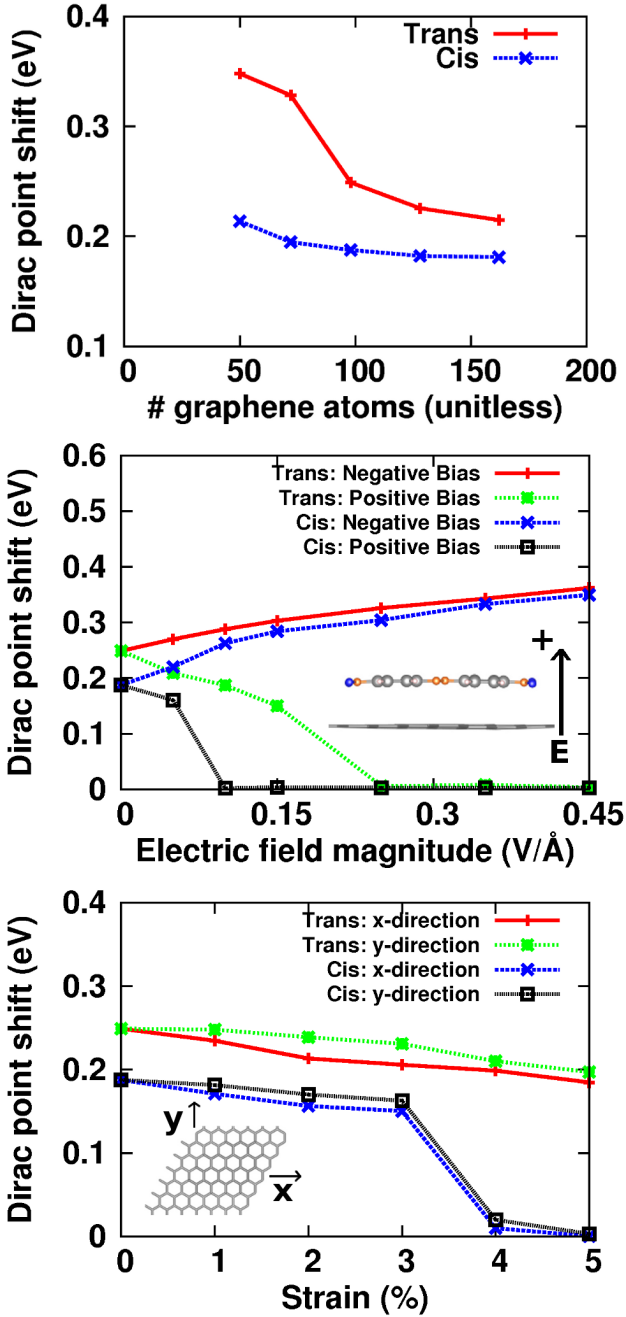}
\caption{\label{efield_strain}(Color online) The Dirac point shift (DPS) using the NO$_2$-AB-NO$_2$ ligands in the $trans$ and $cis$ configuration as a function of a) number of graphene atoms in the supercell, b) an applied electric bias, and c) uniaxial mechanical strain.  Positive and negative electric bias correspond to an electric field pointing from graphene to the molecule and from the molecule to graphene, respectively.  Results for mechanical strain are shown for strain in the x-direction (armchair edge) and the y-direction (zigzag edge).}
\end{figure}

Given that the NO$_2$-AB-NO$_2$ molecule provides the optimal graphene doping differential between isomers, we use this molecule to study the effects of molecular concentration, applied electric bias, and mechanical strain to demonstrate multiple methods of simultaneous doping control.  As shown in Figure \ref{efield_strain}(a), increased molecular concentration leads to a significantly increased DPS for the $trans$ but not the $cis$ configuration.  This is due to the elongated geometry of the $trans$ isomer such that interactions between molecules become significant more quickly at higher concentrations compared to the $cis$ isomer, which has a more vertical adsorption geometry on graphene.  Therefore, we find that the differential doping between isomers and therefore switching potential can be maximized by increasing molecular concentration.  However, extremely dense concentrations will lead to overlap between molecules, lifting NO$_2$ ligands farther from the graphene surface and likely decreasing doping.  Based on our calculations, a 6x6 graphene supercell (72 graphene C atoms per molecule) provides the highest CC ratio of 2.84, corresponding to a DPS of 0.35 and 0.19 for $trans$ and $cis$, respectively.

An applied electric bias is another feasible method to modulate graphene doping.  Figure \ref{efield_strain}(b) plots the DPS for the $trans$ and $cis$ configurations for increasing electric field strength.  Positive or negative bias values correspond to the electric field pointing from the graphene to molecule or from the molecule to graphene, respectively.  As expected, negative electric bias acts in conjunction with the electron-accepting ligands to generate more charge transfer to the molecule and shift the Dirac point to more positive values, indicating greater p-doping.  On the other hand, positive electric bias promotes charge transfer in the opposite direction to the electron-accepting NO$_2$ ligands, thereby shifting the Dirac point closer to the Fermi level and reducing the graphene p-doping.  These effects are seen for both the $trans$ and $cis$ configurations.  For both positive and negative bias, electric fields of large magnitude result in similar DPS for both the $trans$ and $cis$ isomers, indicating that only low magnitudes of bias should be used for switching applications to differentiate between the two configurations.  Charge transfer increases linearly with increased electric field, as expected.  For negative bias, increasing the electric field strength from 0.05 V/$\textup{\AA}$ to 0.45 V/$\textup{\AA}$  increases CT from 0.17$|$e$|$ to 0.48$|$e$|$ for the $trans$ isomer and from 0.10$|$e$|$ to 0.44$|$e$|$ for the $cis$ isomer.  For positive bias, whereas the DPS shows no doping to the level of accuracy in our calculations for higher electric field strength, we find linearly decreasing charge transfer for increasing strength.  Specifically, increasing the electric field strength from 0.05 V/$\textup{\AA}$ to 0.45 V/$\textup{\AA}$  in the positive bias direction decreases CT from 0.09$|$e$|$ to 0.01$|$e$|$ for the $trans$ isomer and from 0.06$|$e$|$ to 0.01$|$e$|$ for the $cis$ isomer.  Interestingly, positive bias never results in a switch from p-doping to n-doping of the graphene monolayer.  Instead, doping levels stay closer to zero from 0.25 to 0.45 V/$\textup{\AA}$ for $trans$ and 0.10 to 0.45 V/$\textup{\AA}$ for $cis$.  This indicates that bias can only be used to modulate levels of p-doping.

Next, we examine the effect of mechanical strain to the graphene monolayer on doping.  Previous studies have shown that applied strain to graphene reduces the Fermi velocity along the strain direction~\cite{pereira2009tight,shishir2009intrinsic} and opens a bandgap for strains greater than 20$\%$~\cite{ribeiro_strain,raoux2009velocity}.  Here, we only examine uniaxial strains up to 5$\%$, relevant to experimental feasibility, and apply uniaxial strains both perpendicular to the armchair edge (x-direction) and zigzag edge (y-direction).  As shown in Figure \ref{efield_strain}(c), we find that uniaxial stress n-dopes the graphene for both strain directions, counteracting the p-doping from the molecule.  Regarding charge transfer, strain along either direction reduces CT to the molecule from 0.13$|$e$|$ (no strain) to 0.09$|$e$|$ (5$\%$ strain) using the $trans$ isomer and from 0.08$|$e$|$ (no strain) to 0.03$|$e$|$ (5$\%$ strain) using the $cis$ isomer.

These results all provide strong evidence that the AB molecule with open shell, electron-accepting ligands can be used for multi-switching applications on graphene monolayers using light, bias, and strain.  As an example that demonstrates the multi-control utility, we consider using the NO$_2$-AB-NO$_2$ derivative on the graphene monolayer.  The $trans$ and $cis$ isomers provide carrier concentrations of 4.16$\times$10$^{12}$ and 2.16$\times$10$^{12}$ cm$^{-2}$ as calculated from the DPS.  If we now apply a negative bias with an electric field strength of 0.05 $V/$$\textup{\AA}$, the carrier concentrations using the $trans$ and $cis$ isomers change to 4.85$\times$10$^{12}$ and 2.94$\times$10$^{12}$ cm$^{-2}$, respectively.  These values indicate four distinct doping states that can be distinguished in experiment and used in a four-state memory or switching device.  Inclusion of mechanical strain would lead to even more switching states.

\begin{center}\textbf{D. Azobenzene Doping of Graphene on Amorphous SiO$_2$ Substrate}\end{center}

\begin{figure}
\includegraphics{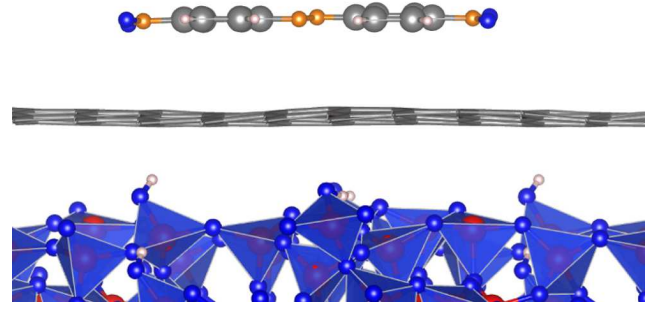}
\caption{\label{sio2}(Color online) Supercell containing the silica substrate (80 SiO$_2$ molecules), 6x6 graphene monolayer (72 graphene C atoms), and NO$_2$-AB-NO$_2$ azobenzene derivative. Small blue spheres are O atoms, small orange spheres are N atoms, small white spheres are H atoms, large grey spheres are C atoms in the benzene rings, and large red spheres are Si atoms.  The thin grey lines correspond to the corrugated graphene monolayer.  Blue tetrahedra represent the four-coordinated amorphous silica structure.  The effects of molecular doping on graphene are examined using 1) fully saturated O bonds at the silica surface, and 2) removing one H atom to leave a dangling bond at the surface to represent p-doped graphene seen in experiment.  See text for details.}
\end{figure}

To this point, we have studied AB molecular doping of a graphene monolayer, however most experiments deposit graphene on an amorphous SiO$_2$ substrate that can intrinsically dope the graphene.  Previous studies of graphene on either $\alpha$- or amorphous SiO$_2$ have shown both p-doping~\cite{shi2009effective,ao2012density,fan2012interaction} due to dangling surface O bonds or n-doping~\cite{shi2009effective,miwa2011doping,romero2008n} due to three-coordinated O atoms in amorphous structures.  The previous experimental paper investigating SO$_3$Na-AB-2CH$_3$ derivatives on graphene found that the SiO$_2$ substrate p-doped graphene prior to depositing the molecules.  To examine whether the SiO$_2$ substrate affects the AB derivative doping behavior, we have examined graphene doping using the NO$_2$-AB-NO$_2$ molecule adsorbed to a 6x6 (72 C atoms) graphene monolayer on an amorphous SiO$_2$ substrate, as shown in Figure \ref{sio2} (computational methods for generating the structure can be found in Section II).

As shown in Figure \ref{sio2}, the graphene monolayer deposited on the amorphous SiO$_2$ substrate becomes corrugated and roughly follows the undulations of the SiO$_2$ surface.  The average graphene-SiO$_2$ vertical distance is 3.64 $\textup{\AA}$ and the average formation energy of the surface is 5.7 meV/$\textup{\AA}$$^2$, matching previous first-principles~\cite{miwa2011doping,romero2008n} and experimental results~\cite{ishigami2007atomic}.  The average level of corrugation in the oxide, measured as the vertical distance between the lowest and highest SiO$_2$ surface atoms, is 1.51 $\textup{\AA}$.  The average graphene corrugation of 0.43 $\textup{\AA}$ is much less than the SiO$_2$ surface, which can be attributed to the large energy cost to create vertical disruptions in the graphene lattice.

Table \ref{table_sio2} reports the Dirac point shift (DPS) for graphene on the SiO$_2$ substrate for three cases: 1) no AB molecule, 2) $trans$ NO$_2$-AB-NO$_2$ molecule, and 3) $cis$ NO$_2$-AB-NO$_2$ molecule. With no molecule, we find that the SiO$_2$ substrate with saturated surface bonds does not dope the graphene, leading to no DPS.  Previous first-principles studies have shown that SiO$_2$ with saturated surface bonds can n-dope graphene due to three-coordinated O atoms~\cite{miwa2011doping}, however our samples show a distribution of three- and five-coordinated O atoms that leads to no doping.  However, if we remove an H atom (white atom in Figure \ref{sio2}) from the surface to reveal an O dangling bond, the SiO$_2$ substrate p-dopes the graphene, leading to a 0.66 DPS and hole concentration of 2.9$\times$10$^{13}$ cm$^{-2}$.

When we add the NO$_2$-AB-NO$_2$ molecule, we find that the doping levels slightly decrease for both $trans$ and $cis$ when using a saturated SiO$_2$ substrate compared to the graphene monolayer alone.  This is likely due to the corrugation of the graphene layer due to the SiO$_2$ substrate.  As seen in Figure \ref{sio2}, the geometry of the $trans$ AB molecule does not significantly change to follow the corrugations of the graphene.  Therefore, the average distance between the graphene and molecule increases from 3.16 $\textup{\AA}$ for the monolayer alone to an average of 3.23 $\textup{\AA}$ when using a SiO$_2$ substrate. In particular, the NO$_2$-graphene distance, critical for doping, increases from 3.11 to 3.18 $\textup{\AA}$.  Nevertheless, a substantial difference persists between the $trans$ and $cis$ configurations, indicating that switching behavior is robust even when graphene is supported on a substrate with no dangling bonds.  The carrier concentrations for each DPS on the saturated substrate correspond to 4.45$\times$10$^{12}$ and 1.83$\times$10$^{12}$ cm$^{-2}$ for $trans$ and $cis$, respectively, giving a CC ratio of 2.43.

\begin{table}
\renewcommand{\tabcolsep}{0.30cm}
\small
\caption{\label{table_sio2} Comparison of the Dirac point shift (DPS) (eV) and carrier concentration (CC) ($\times$10$^{13}$ cm$^{-2}$) in graphene on a silica substrate with and without the NO$_2$-AB-NO$_2$ molecule.  Doping has been investigated for the graphene system on a silica substrate with saturated bonds (saturated) and with one dangling O bond (dangling bond).  A 6x6 graphene supercell has been used for these calculations.  See text for details.}
\resizebox{\columnwidth}{!}{%
\begin{tabular}{l*{15}{c}r}
isomer & monolayer & & saturated & & dangling & \\
\hline
& DPS & CC & DPS & CC & DPS & CC \\
\hline
no molecule & 0.00 & 0.00 & 0.00 & 0.00 & 0.66 & 2.9 \\
$\textit{trans}$ & 0.33 & 0.72 & 0.26 & 0.45 & 0.70 & 3.26 \\
$\textit{cis}$ & 0.19 & 0.25 & 0.17 & 0.18 & 0.67 & 3.02 \\
\hline
\end{tabular}}
\end{table}

This situation drastically changes when we remove one H surface atom from the SiO$_2$ substrate.  In this case, the SiO$_2$ substrate p-dopes the graphene, leading to a 0.66 eV DPS, and the molecule then has little effect on further doping of the graphene monolayer.  The $trans$ isomer only increases this doping to a 0.70 eV DPS, and the $cis$ isomer to a 0.67 eV DPS.  These correspond to carrier concentrations of 3.26$\times$10$^{13}$ and 3.02$\times$10$^{13}$ cm$^{-2}$ and a CC ratio of only 1.08, significantly reducing the potential for doping differentials between isomers.  This indicates that the O atoms in the SiO$_2$ substrate act as much stronger electron acceptors compared to the ligands in the AB molecule.  These results emphasize the importance in reducing the p-doping from the SiO$_2$ substrate to maximize the switching behavior of the AB molecule.  

\begin{center}\textbf{IV. CONCLUSIONS}\end{center}
\label{Conclusions}

We have presented comprehensive first-principles data regarding graphene doping using the azobenzene (AB) molecule functionalized with both electron-accepting and -donating ligands.  We find that open shell ligands are crucial for substantial charge transfer between the graphene and molecule that leads to a shift in the Dirac point and induced carrier concentration.  Across all ligands tested, we find that SO$_3$ and NO$_2$ are the best candidates for p-doping as indicated by large charge transfer from the graphene to the molecule and a significant shift in graphene's Dirac point to higher energies compared to its position in pristine graphene.  By comparing the change in graphene carrier concentration across ligand types, we find that the NO$_2$-AB-NO$_2$ derivative provides the best doping differential between $trans$ and $cis$ isomers with a carrier concentration ratio of 1.92.  Our analysis of charge density difference isosurfaces and partial density of states indicate that changes in doping between isomers is due to one benzene ring lifting farther from the surface in the $cis$ configuration, reducing its ability to dope graphene.  Maximum differences between $trans$ and $cis$ doping can be achieved using a moderate electron-acceptor, such as the NO$_2$ ligand, compared to SO$_3$, which is a strong enough acceptor to significantly dope graphene even when lifted farther from the surface in the $cis$ isomer.  This reduces its resulting doping differential between isomers compared to the NO$_2$ ligand.  These results indicate that a careful choice of ligand is crucial to maximize swtiching effects, in that the ligand must be a strong enough acceptor to dope the graphene at all, but weak enough such that the change in distance from the graphene between the $trans$ and $cis$ isomers significantly decreases its doping potential.  Relevant to experiment, we have found that the AB derivatives can significantly dope graphene supported on a silica substrate as long as dangling bonds from O atoms in the silica have been saturated.

In addition to demonstrating the potential for graphene doping through photoisomerization, we have also shown that doping levels due to either isomer can be futher modulated through the application of an electrical bias or mechanical strain to the graphene monolayer.  These results are the first demonstration of such multi-control graphene doping using an organic molecule.  A positive bias, applied from the graphene to the molecule, linearly reduces charge transfer and subsequent doping effects for both $trans$ and $cis$ isomers.  Increasing positive bias beyond 0.10 V/$\textup{\AA}$ for $cis$ and 0.25 V/$\textup{\AA}$ for $trans$ completely eliminates any p-doping of graphene, but never reverses to an n-type doping.  Negative bias linearly enhances charge transfer and to a greater degree for the $cis$ compared to $trans$, such that doping levels are similar for both at high bias.  Using low positive or negative bias on the order of $|$0.05$|$ to $|$0.10$|$ V/$\textup{\AA}$, we have demonstrated the possiblity for multi-level graphene doping by combining photoisomerization with the applied electric field.  We also show a similar multi-control doping mechanim using mechanical strain of the graphene monolayer, which decreases the p-doping induced by the AB molecule.  These valuable findings should encourage experimental confirmation and the fabrication of such a multi-switch device.

\begin{acknowledgements}
This work is supported by the Department of Energy (DOE), Office of Basic Energy Sciences (BES), under Contract No. DE-FG02-02ER45995.
\end{acknowledgements}





%

\end{document}